\documentstyle[12pt,twoside]{article}
\newcommand{\be}{\begin{equation}}
\newcommand{\ee}{\end{equation}}
\begin{document}
 

\title{\small \rm \begin{flushright} \small{hep-ph/9607349}\\
\small{RAL-96-051}\\
\small{TAUP-2352-96}\\
\small{WIS-96/XXXXXXXXXX} \end{flushright} \vspace{2cm}
\LARGE \bf Final State Interactions, Resonances and CP Violation 
in D and B Exclusive Decays\vspace{0.8cm}}
\author{Frank E. Close\thanks{E-mail : fec@v2.rl.ac.uk}\thanks{Work
supported in part by the European Community Human Capital Mobility program
``Eurodafne", Contract CHRX-CT92-0026.} \\
{\small \em Particle Theory, Rutherford-Appleton Laboratory, Chilton,
Didcot OX11 0QX, UK} \\ \\
Harry J. Lipkin \thanks{E-mail : ftlipkin@ weizmann.weizmann.ac.il} 
\thanks{Work supported in part by the German-Israeli Foundation for Scientific 
Research and Development (GIF).} \\
{\small \em
Department of Particle Physics,Weizmann Institute of Science,
Rehovot 76100, Israel}
\\
\centerline{and}
\\
{\small \em School of Physics and Astronomy,
Tel Aviv University,Tel Aviv, Israel}
    \\  \\}
\date{July 12 1996 \vspace{1.5cm}}

\begin{center}
\maketitle

\begin{abstract}
Hadron resonances affect nonexotic $D^o$ decays but not $B$ decays
which are far from the resonance region. 
We obtain new information from {\bf exclusive} decays and show
that interference between colour favoured and
colour suppressed diagrams is 
{\bf constructive} in $B$ and (some) $D$ decays
in contrast to the inclusive decays where a net Pauli destructive interference
is claimed. We suggest that a systematic study of $B$ decay final states
containing excited mesons such as $a_1 D$ or $\pi D^{**}$ show opposite
behaviour for $B^o$ and $B^+$ relative to the ground state channels.
We give inequalities among final state branching ratios for several
$B$ and $D$ channels. Decays into $\pi K^{(*)}$ and $\rho K^{(*)}$
appear to show different behaviours.
\end{abstract}
\end{center}
 
\newpage
 
\section{Introduction }
 
The role of final state interactions and possible hadron resonances in weak
decays of heavy flavour hadrons remains to be understood, particularly in
view of the presence of known meson resonances in the vicinity of the D
mass\cite{cll,ves}. In the present paper we 
look for clues to possible resonance effects in the systematics of decay
branching ratios.
 
We first note that the final states of Cabibbo-favored 
dominant decay modes of charged $D$ and charged $B$ decays have the
exotic isospin quantum number I = 3/2 which does not exist for the
quark-antiquark system. There are therefore no meson resonances contributing
to these final states. The corresponding neutral $D$ and neutral $B$ decays
have final states with nonexotic isospin I=1/2 as well as the exotic I=3/2
and can have resonance contributions. It is therefore of interest to look for
differences in the systematics of charged and neutral decays. Note that the 
isospin couplings of corresponding $D$ and $B$ decay diagrams are identical, 
since they differ only in interchanging the two isoscalar transitions 
$c \rightarrow s$ and $b \rightarrow c$. 

One immediately
finds that the neutral and charged $D$ lifetimes are different but that neither
the semileptonic partial widths for $D$ decays nor the total $B$ widths show
such a difference. This suggests
that meson resonances are responsible for speeding up the hadronic decays 
of the $D^o$ 
while such effects can be absent at the $B$ mass which is far above the
resonance region. We also note that the theoretical basis of the
$\Delta I = 1/2$ rule in kaon decays still is not fully understood
and that the $I=2$ state of the two-pion system which is suppressed by
$\Delta I = 1/2$ also has exotic isospin. There is also the unexplained
enhancement of nonleptonic weak decays relative to semileptonics.
There seems to be a general enhancement of weak decays into mesonic final
states having nonexotic quantum numbers relative to those having exotic
quantum numbers and also relative to semileptonic decays.
 
Before rushing to explain everything with final state resonances, we note
that other explanations have been proposed for the $D$ lifetime
difference\cite{bigi}.
The W-exchange weak diagram, in which a W boson is exchanged between the
initial quark-antiquark pair exists only for the neutral initial states and
provides an extra I=1/2 contribution. There are arguments suggesting a mass
dependence for this diagram which suppresses this contribution at the $B$
mass. There have also been arguments pointing out that the two possible colour
couplings, generally called colour favored and colour suppressed, lead to the
same final state in charged decays; e.g. $K^+\pi^o$  and lead to different
final states; e.g. $K^+\pi^-$ and $K^o\pi^o$ in neutral decays. Thus 
in the charged decays the two
amplitudes can interfere destructively in what has been called a ``Pauli"
effect thereby suppressing them relative to the neutral
decays.
 
The explanations using these weak diagrams are still subject to intense
controversy\cite{alt}. The sums over final states to obtain decay rates use
a quark basis and neglect all details of hadron spectroscopy such as the nature
of the low-lying hadronic final states which have the largest phase space. This
phase space factor dominates the $K_L$ - $K_S$ lifetime difference and there 
are suggestions that similar effects determine the lifetime difference of the
$B_s$ eigenstates\cite{CPeven,Blifes}. 
Furthermore the mass dependence and the sign of the 
``Pauli" interference have been questioned. In order to obtain additional input
from experiment to help resolve these questions we investigate what can be 
learned from
exclusive decay modes, where there are a wealth of data constantly becoming
available and where new systematics beyond what is available in inclusive
decays can give new clues to the underlying mechanisms.
 
In sections 2 and 3 we present a preliminary detailed analysis 
of the decays to the lowest-lying exclusive quasi-two-body final states which 
shows that the nonexotic I=1/2 contribution is enhanced in D decays 
(section 2) but not in B decays  (section 3) 
as expected either from a resonance argument or from W exchange
with some mass factor. This  enhancement is seen in the nonexotic exclusive D
decays as well as in the lifetime difference. The systematics of $B$ decays
are in marked contrast to this; not only is there no enhancement of the 
non-exotic exclusive $B$ decays but one finds that the exotic transitions 
are enhanced relative to the nonexotics! This indication  for $constructive$
interference
between colour favored and colour suppressed contributions to the exotic final
states (rather than destructive as suggested by the Pauli
argument) is also supported by our analysis of the exclusive D decays
(section 4).
 
This raises the interesting point\cite{Grossman} that this exotic enhancement
of the exclusive B decays to low lying final states (e.g. $\pi D$)
would contradict the observed absence of
enhancement in the inclusive decays indicated by the approximate equality of
the charged and neutral lifetimes. This suggests that there are form factor
effects depending upon final state wave functions for decays into
higher states (e.g. $a_1D, \pi D^{**}$) which reverse the relative
signs of the colour favored and colour suppressed contributions to different
final states. We comment on this in section 5.
  
Our analysis uses the language of the weak parton model but is based on a 
much more general flavour-topology formulation which includes all final state 
interaction effects\cite{PKEKETA}. 
A heavy-flavoured quark and a light-flavoured antiquark are
assumed to enter a black box from which two final $q \bar q$ pairs emerge. The
initial nonstrange quark line is assumed to travel in all possible complicated 
paths going forward and backward in time and emitting and absorbing gluons 
until it either disappears in the box by interaction in a weak vertex (W) or 
emerges from the box as a constituent of the final charmed or strange
meson (T) or as a constituent of the final nonstrange meson (S). Since all 
strong interactions are assumed to conserve isospin, the additional $q \bar q$ 
pair created by gluons within the box in the W topology must be isoscalar, and
the contributions to the T and S diagrams must be independent of the isospin 
of the initial nonstrange quark which travels through the box undergoing only
isospin-invariant strong interactions. The familiar parton model coloour 
favoured, colour suppressed and $W$ diagrams are a particular example of this 
more general topological classification.
 
\section{A Systematic Enhancement of Nonexotic $D^o$ Decays }
 
As a first step it is interesting to compare the experimental
branching ratios for $D^+$ to exotic
I=3/2 states with the corresponding branching ratios of $D^o$ to charged
and neutral final states which are mixtures of I=3/2 and I=1/2 and
therefore have a non-exotic component. 
Note that the exotic I=3/2 $D^+$ amplitudes have both
colour-favored and-colour suppressed amplitudes, while the
$D^o$ amplitudes to charged final states are purely colour-favored and
the $D^o$ amplitudes to neutral final states are purely colour suppressed.
To enable focusing on systematics in
comparing decays to different final states with different wave functions and
form factors, we note that the color-favored tree amplitudes should have a 
point-like form factor e.g. the wave function at the origin, for the charged  
nonstrange meson and a hadronic form factor for the strange meson nearly the 
same as that for the semileptonic decay to the same strange meson. We therefore
use this semileptonic branching ratio for normalization and express the errors
separately for the hadronic numerator and the semileptonic denominator.
 
$\bullet$ For the exotic $
\pi^+\bar K^o
$ final state we obtain\cite{pdg94}:
$$
{{br(\pi^+\bar K^o)}\over{br(\nu e^+\bar K^o)}}=
0.42 \pm 0.04
\pm 13\%
\eqno(1a) $$
in comparison with the nonexotic colour-favored and colour-suppressed
$$
{{br(\pi^+K^-)}\over{br(\nu e^+K^-)}}=
1.05 \pm 0.04
\pm 6\% ;
~ ~ ~ ~ ~ ~
{{br(\pi^o\bar K^o)}\over{br(\nu e^+K^-)}}=
0.54 \pm 0.07
\pm 6\%
\eqno(1b) $$
$\bullet$For the exotic $
\rho^+\bar K^o
$ final state we obtain:
$$
{{br(\rho^+\bar K^o)}\over{br(\nu e^+\bar K^o)}}=
1.0  \pm 0.37
\pm 13\%
\eqno(2a) $$
in comparison with the nonexotic colour-favored and colour-suppressed
$$
{{br(\rho^+K^-)}\over{br(\nu e^+K^-)}}=
2.74 \pm 0.34
\pm 6\%
~ ~ ~ ~ ~ ~
{{br(\rho^o\bar K^o)}\over{br(\nu e^+K^-)}}=
0.29 \pm 0.05
\pm 6\%
\eqno(2b) $$
$$
{{br(\omega\bar K^o)}\over{br(\nu e^+K^-)}}=
0.53 \pm 0.11
\pm 6\%
\eqno(2c) $$
$\bullet$For the exotic $
a_1(1260)^+\bar K^o
$ final state we obtain:
$$
{{br(a_1(1260)^+\bar K^o)}\over{br(\nu e^+\bar K^o)}}=
1.23 \pm 0.25
\pm 13\%
\eqno(3a) $$
in comparison with the nonexotic colour-favored and colour-suppressed
$$
{{br(a_1(1260)^+K^-)}\over{br(\nu e^+K^-)}}=
2.08 \pm 0.31
\pm 6\%
~ ~ ~ ~ ~ ~
{{br(a_1(1260)^o\bar K^o)}\over{br(\nu e^+K^-)}} <
0.13
\eqno(3b) $$
$\bullet$For the exotic $
\pi^+\bar K^*(892)^o
$ final state we obtain:
$$
{{br(\pi^+\bar K^*(892)^o)}\over{br(\nu e^+\bar K^*(892)^o)}}=
0.47 \pm 0.09
\pm 10\%
\eqno(4a) $$
in comparison with the nonexotic colour-favored and colour-suppressed
$$
{{br(\pi^+K^*(892)^-)}\over{br(\nu e^+K^*(892)^-)}}=
3.8 \pm 0.46
\pm 23\%
~ ~ ~ ~ ~ ~
{{br(\pi^o\bar K^*(892)^o)}\over{br(\nu e^+K^*(892)^-)}}=
2.30 \pm 0.23
\pm 23\%
\eqno(4b) $$
$\bullet$ For the exotic $
\rho^+\bar K^*(892)^o
$ final state we obtain:
$$
{{br(\rho^+\bar K^*(892)^o)}\over{br(\nu e^+K^*(892)^o)}}=
0.43 \pm 0.28
\pm 10\%
\eqno(5a) $$
in comparison with the nonexotic colour-favored and colour-suppressed
$$
{{br(\rho^+K^*(892)^-)}\over{br(\nu e^+K^*(892)^-)}}=
4.54 \pm 0.18
\pm 23\%
~ ~ ~ ~ ~ ~
{{br(\rho^o\bar K^*(892)^o)}\over{br(\nu e^+K^*(892)^-)}}=
1.23 \pm 0.31
\pm 23\%
\eqno(5b) $$
 
{\it There appears to be a systematic enhancement of the nonexotic
$D^o$ decays relative to the exotic $D^+$ decays described by exactly
the same leading diagrams}.

 \section{Exotic-Nonexotic Systematics of B  Decays }
 
For B decays treated by analogy with the foregoing D decays, we find a radically
different systematics. First we list the modes by analogy with the $D$ decays
in section 2. 

$\bullet$For the exotic $\pi^+\bar D^o$ final state we obtain\cite{pdg94}:
$$
{{br(\pi^+\bar D^o)}\over{br(\nu e^+\bar D^o)}}=
0.33 \pm 0.03
\pm 44\%
\eqno(B1a) $$
in comparison with the nonexotic colour-favored and colour-suppressed
$$
{{br(\pi^+D^-)}\over{br(\nu e^+D^-)}}=
0.16 \pm 0.02
\pm 26\% ;
~ ~ ~ ~ ~ ~
{{br(\pi^o\bar D^o)}\over{br(\nu e^+D^-)}}
\leq 0.02
\pm 26\%
\leq 0.025
\eqno(B1b) $$

$\bullet$For the exotic $
\rho^+\bar D^o
$ final state we obtain:
$$
{{br(\rho^+\bar D^o)}\over{br(\nu e^+\bar D^o)}}=
0.83 \pm 0.11
\pm 44\%
\eqno(B2a) $$
in comparison with the nonexotic colour-favored and colour-suppressed
$$
{{br(\rho^+D^-)}\over{br(\nu e^+D^-)}}=
0.41 \pm 0.07
\pm 26\%
~ ~ ~ ~ ~ ~
{{br(\rho^o\bar D^o)}\over{br(\nu e^+D^-)}}
\leq 0.04
\eqno(B2b) $$
$$
{{br(\omega\bar D^o)}\over{br(\nu e^+D^-)}}
\leq 0.05(?)
\eqno(B2c) $$
$\bullet$ For the exotic $
a_1(1260)^+\bar D^o
$ final state we obtain:
$$
{{br(a_1(1260)^+\bar D^o)}\over{br(\nu e^+\bar D^o)}}=
0.31 \pm 0.25
\pm 44\%
\eqno(B3a) $$
in comparison with the nonexotic colour-favored and colour-suppressed
$$
{{br(a_1(1260)^+D^-)}\over{br(\nu e^+D^-)}}=
0.31 \pm 0.17
\pm 26\%
~ ~ ~ ~ ~ ~
{{br(a_1(1260)^o\bar D^o)}\over{br(\nu e^+D^-)}} <
???
\eqno(B3b) $$
$\bullet$For the exotic $
\pi^+\bar D^{*o}
$ final state we obtain:
$$
{{br(\pi^+\bar D^{*o})}\over{br(\nu e^+\bar D^{*o})}}=
0.08 \pm 0.01
\pm 33\%
\eqno(B4a) $$
in comparison with the nonexotic colour-favored and colour-suppressed
$$
{{br(\pi^+D^{*-})}\over{br(\nu e^+D^{*-})}}=
0.06 \pm 0.01
\pm 11\%
~ ~ ~ ~ ~ ~
{{br(\pi^o\bar D^{*o})}\over{br(\nu e^+D^{*-})}}
< 0.03
\eqno(B4b) $$
$\bullet$For the exotic $
\rho^+\bar D^{*o}
$ final state we obtain:
$$
{{br(\rho^+\bar D^{*o})}\over{br(\nu e^+\bar D^{*o})}}=
0.23 \pm 0.05
\pm 33\%
\eqno(B5a) $$
in comparison with the nonexotic colour-favored and colour-suppressed
$$
{{br(\rho^+D^{*-})}\over{br(\nu e^+D^{*-})}}=
0.17 \pm 0.03
\pm 11\%
~ ~ ~ ~ ~ ~
{{br(\rho^o\bar D^{*o})}\over{br(\nu e^+D^{*-})}}
< 0.03
\eqno(B5b) $$
 
{\it In contrast to D decays where there
appears to be a systematic enhancement of the nonexotic
$D^o$ decays relative to the exotic $D^+$ decays described by exactly
the same leading diagrams, here we find the exotic modes tend to be slightly
larger than the non-exotic. There is also a drastic suppression of the
non-exotic colour suppressed}.

\section{A Weak Diagram Analysis}
 
There are two different approaches to the nonexotic enhancement present in 
$D$ decays. One is to attribute it to strong final state interactions in 
channels having resonances\cite{buccella,kamal}.
However there are also attempts to explain it via weak interaction diagrams
without taking final state interactions into account\cite{bigi,alt}. 
Both approaches explain the absence of such enhancement in $B$ 
decays, but the ``reverse enhancement" of exotics noted above in low-lying 
exclusive $B$ decays has not previously been discussed.

In the weak interaction approach there are three types of contributions to 
these Cabibbo-favored $D$ decays in the standard model:
(1) a colour-favored tree diagram; (2) a colour-suppressed tree diagram; (3)
a W-exchange diagram. We use this formalism but interpret results using the
general flavour-topology approach described in the introduction.
Both tree diagrams contribute to the exotic channels
but the W-exchange diagram does not contribute since it goes via an intermediate
state containing only a single $q \bar q$ pair. In the nonexotic channels there
are W-exchange contributions and either a colour-favored or colour-suppressed
diagram, but not both. The argument then goes that there is an enhancement in
the non-exotic channels due to the W-exchange diagram, and there is a
suppression in the exotic channels due to so-called ``Pauli" interference
between the colour-favored and colour-suppressed diagrams, which is claimed to
be always destructive\cite{bigi}.  This claim, however, is based on general 
arguments that
apply to inclusive $D$ decays and whether it is correct and whether it applies
universally to all exclusive channels is open to question and to experimental
tests. 
 
As a first test to see how this can work, we express the the amplitudes for
the vector-pseudoscalar decay modes in terms of these three contributions
denoted by T, S and W respectively for colour-favored tree, colour-suppressed
tree and W exchange. Note that any contribution due to final state interactions
which go via an intermediate $q \bar q$ state is pure $I=1/2$ and has exactly 
the same couplings to all decays as the $W$ contribution\cite{PKEKETA}. 
Thus our analysis 
below is completely general and includes these final state and resonance 
contributions. However we cannot determine at this stage how much of the $W$
contribution is due to weak W exchange and how much is due to strong final state
enhancements; e.g. resonances.
 
$$A_D(\rho^+\bar K^o)  =  T + S , ~ ~ ~ ; ~ ~ ~ 
A_B(\rho^+\bar D^o)  =  T_B + S_B , \eqno(6a) $$
$$A_D(\rho^+K^-) = T + W          ,~ ~ ~ ; ~ ~ ~ 
A_B(\rho^+D^-) = T_B + W_B          ,\eqno(6b) $$
$$A_D(V_u\bar K^o) = S             ,~ ~ ~ ; ~ ~ ~
A_B(V_u\bar D^o) = S_B             ,\eqno(6c) $$
$$A_D(V_d\bar K^o) = W             ,~ ~ ~ ; ~ ~ ~
A_B(V_d\bar D^o) = W_B             ,\eqno(6d) $$
$$A_D(\rho^o\bar K^o) = { 1 \over {\sqrt 2}}\cdot (S - W) , ~ ~ ~ ; ~ ~ ~ 
A_B(\rho^o\bar D^o) = { 1 \over {\sqrt 2}}\cdot (S_B - W_B) , \eqno(6e) $$
$$A_D(\omega\bar K^o) = { 1 \over {\sqrt 2}}\cdot (S + W) , ~ ~ ~ ; ~ ~ ~ 
A_B(\omega\bar D^o) = { 1 \over {\sqrt 2}}\cdot (S_B + W_B) , \eqno(6f) $$
$$A_D(\phi\bar K^o) = \xi W ~ ~ ~ ; ~ ~ ~ 
A_B(\phi\bar D^o) = \xi W_B \eqno(6g) $$
where we have used the notation $V_u$ and $V_d$ for the $u \bar u$ and
$d \bar d$ vector meson states and noted that the
$ \rho^o $ and $\omega$ are equal mixtures of these two states with opposite
relative phase. $\xi$ is a flavour-SU(3)-breaking parameter expressing the
suppression of creating strange quark pairs from the vacuum. We do not use
$A(\phi\bar K^o)$ in our subsequent analysis since it adds one piece of data
with one free parameter. The parameter $\xi$ can be determined from experiment 
if desired just to check internal consistency.
 
At this stage we have four experimental quantities expressed in terms of three
complex amplitudes and therefore in terms of five parameters. But we can get a
qualitative picture if we simply assume that all amplitudes are relatively
real. We now have four quantities overdetermining three parameters and
we can see whether these can fit the data. 
\subsection {Application to $D \rightarrow \rho K$ Decays}

We find that the data
can indeed be fit by setting
$$T = 0.81 \eqno(7a) $$
$$W = 0.87        \eqno(7b) $$
$$S = 0.17        \eqno(7c) $$
where we have normalized the amplitudes so that
$$
{{br(\rho^+\bar K^o)}\over{br(\nu e^+\bar K^o)}}=
1.0  \pm 0.37
\pm 13\% = (T + S)^2 = 0.96
\eqno(8a) $$
$$
{{br(\rho^+K^-)}\over{br(\nu e^+K^-)}}=
2.74 \pm 0.34 \pm 6\% = (T + W)^2 = 2.82
\eqno(8b) $$
$$
{{br(\rho^o\bar K^o)}\over{br(\nu e^+K^-)}}=
0.29 \pm 0.05
\pm 6\% = (1/2)\cdot (W - S)^2 = 0.25
\eqno(8c) $$
$$
{{br(\omega\bar K^o)}\over{br(\nu e^+K^-)}}=
0.53 \pm 0.11
\pm 6\% = (1/2)\cdot (W + S)^2 = 0.54
\eqno(8d) $$
 
We note the following qualitative feature of this fit.
The colour-favored tree and the W-exchange amplitudes are roughly equal and the
colour-suppressed tree amplitude is much smaller. The interference between the
colour-favored and the colour-suppressed tree amplitudes is constructive
in the exotic $\rho^+\bar K^o$ decay, in contradiction with the ``Pauli effect"
which predicts destructive interference.
 
The basic physics in this qualitative argument lies in relative phases
determined by the isospinology and the experimental result that
$$|A(\rho^o\bar K^o)| = { 1 \over {\sqrt 2}}\cdot |(S - W)| <
|A(\omega\bar K^o)| = { 1 \over {\sqrt 2}}\cdot| (S + W)|  \eqno(9) $$
This tells us that the interference between the $S$ and $W$ amplitudes for
$A(\rho^o\bar K^o)$ must be destructive. Thus with the phase convention
chosen for eqs. (6) the experimental result (9)
requires positive relative phase for the S and W amplitudes.
 
The tree amplitudes for the $\rho-K$ final states have the same phase when
the flavour of the spectator quark is changed since the change of the
spectator quark is an isospin raising or lowering operator.
 
The W-exchange
amplitudes have the opposite phase for the charged and neutral $\rho$ decays
because they arise from an $I=1/2$ $q \bar q$ state. The creation of an
additional nonstrange pair by gluons conserves isospin, and the isospin
Clebsches for coupling the I=1 $\rho$ with the I=1/2 kaon to a total isospin
of I=1/2 have opposite relative phase for the charged and neutral modes.
The physics of the negative Clebsch is simple. When a spin of 1/2 is coupled
to a spin of 1, the two spins can be either parallel or antiparallel. When
they are parallel the total spin is 3/2. To make a total spin of 1/2 the
individual spins must be antiparallel. This gives the negative phase.
 
These arguments determine uniquely the relative phases in eqs. (6).
The T-S interference in $\rho^+\bar K^o $ and the T-W interference in
$\rho^+K^-$ are thus required to be the same; i.e. either both constructive or
both destructive. Note that the $\rho-\omega$ mixing of $u \bar u$ and
$d \bar d$ states plays a crucial role in this analysis and allows the relative
phase of the $W$ and $S$ amplitudes to be determined by the experimental
inequality (9).
 
These features tend to support the picture of resonance enhancement in the
nonexotic channels. In this picture the $W$ amplitude has a contribution from
final-state rescattering, which has the same topology as the W exchange. Thus
the prominent 
W amplitude may be largely due to final state resonance scattering.
 
The relative phase of the T and S amplitudes can in principle be calculated
from the standard model and hadron wave functions for the mesons. This is a
complicated calculation involving point-like and hadronic form factors for
the $\rho$ and $K$ mesons and colour and spin recouplings. These complications
are avoided in the calculations for inclusive processes where the arguments
for the ``Pauli relative phase" may be valid. The present exclusive process
can certainly not be described as a simple Pauli effect. We have avoided this
calculation by using the
experimental inequality (9) as input. A correct calculation of the relative
phase would predict this inequality and hopefully would agree with experiment.
 
\subsection{Application to $B \rightarrow X D$ Decays}
 
In the $B \rightarrow X D$ Decays where $X$ denotes any isovector meson, 
detailed analyses analogous to those above for $D$ decays are presently masked 
by the large error bars, but the improvements anticipated from B-factories and 
elsewhere should enable sharper quantification soon. However, it is already
clear from the small upper bounds on the decays into two neutral particles that
both the $W_B$ and $S_B$ amplitudes are small. We can therefore analyze the data
using expressions to lowest order in these small amplitudes. It is convenient to
define:

$$\Gamma^{+o} \equiv  |A_B(X^+\bar D^o)|^2  =  |T_B + S_B|^2 \approx
T_B^2 + 2T_B \cdot S_B, \eqno(10a) $$
$$\Gamma^{+-} \equiv  |A_B(X^+D^-)|^2 = |T_B + W_B|^2  \approx
T_B^2 + 2T_B \cdot W_B
          ,\eqno(10b) $$
$$\Gamma^{oo}  \equiv  |A_B(X^o\bar D^o)|^2 = { 1 \over { 2}}\cdot 
|(S_B - W_B)|^2 . \eqno(10c) $$
Then to lowest order in the small amplitudes,
$${ 1 \over {4}}\cdot {{|\Gamma^{+o} - \Gamma^{+-}|^2}\over
{\Gamma^{+o} + \Gamma^{+-}}}\approx   
{{|T_B \cdot (S_B - W_B) |^2 }\over{2 T_B^2 }}= \Gamma^{oo} \cos^2 \theta  
\eqno(11a) $$
where 
$$ \cos \theta  \equiv 
{{T_B \cdot (S_B - W_B) }\over{|T_B| |S_B - W_B| }}
\eqno(11b) $$
The present data show only upper bounds for $\Gamma^{oo} $ which satisfy these 
relations for all the $X D$ states given above in eqs. (B1-B5).  

However, these data already reveal the interesting and surprising systematics
that {\it for $B$ decays  the exotic branching ratios are consistently larger 
than the nonexotic} and that this difference comes from an interference term
between the $T$ and $W-S$. The direct terms proportional to squares of these
small amplitudes are seen from the data to be below the presently measured 
upper limits. 
Although in principle the relation (11) does not specify which of the two
small amplitudes dominates in the interference term, it seems hardly likely that
the $W$ amplitude should have opposite phase in $D$ and $B$ decays 
(i.e. that it would be constructive in $D$ and destructive in $B$ decays).
One rather assumes
that just as in D decays the colour-suppressed amplitude is
small, but interferes {\bf constructively} with the colour-favored amplitude.
This completely disagrees with the conventional weak diagram folklore,
where this ``Pauli" interference is predicted to be destructive in both
cases.

Thus we see that in contrast to the D decays where
 
$$W_D \geq T_D > S_D \eqno(12a) $$
the B system shows
$$T_B > S_B >> W_B. \eqno(12b) $$
 
It is interesting to note that these results are at least qualitatively
in accord with expectations from the presence of direct channel
resonance enhancements. However it is not surprising
that the W-exchange goes away. The weak interaction calculators say that
this results naturally from mass factors in the diagram.

\section{Comparison of $B$ and $D$ Decays}

For a ball-park estimate set T=3, W=3, and S=1 for D decays and the
same with W=0 for B decays. These values are not normalized;
only ratios are relevant. We then obtain for the ratios:
 
$$
\Gamma_D^{+o}/\Gamma_D^{+-}/\Gamma_D^{oo} =  4:9: \frac{1}{2}
\eqno(13a)
$$
for D decays and
$$
\Gamma_B^{+o}/\Gamma_B^{+-}/\Gamma_B^{oo} = 16:9: \frac{1}{2}
\eqno(13b)
$$
for B decays.
 
This is clearly oversimplified, since there is no reason to believe that
all amplitudes are relatively real. But the qualitative prediction that
the exotic branching ratios are systematically lower than the decays into
two charged particles by a factor of two in D decays and systematically
higher by a factor of two for B decays is impressive.
 
There is also the qualitative feature that a small colour-suppressed
amplitude can give a significant enhancement to the exotic amplitude by
constructive interference, while its direct contribution to the neutral
decay is down by an order of magnitude.
 
      An interesting contrast between B and D decay systematics in the above
analysis has been pointed out by Yuval Grossman\cite{Grossman}.
In D decays the
systematic enhancement of decay rates in nonexotic channels is seen in total
decay rates; the lifetime of the neutral D being shorter than the charged D.
In the B system there is no such overall enhancement; the lifetimes are equal
at the level of experimental errors. Thus the ``reverse enhancement"
observed in the $B$ decays and
which are as large as factors of two favoring the charged modes
cannot be general. It is very likely that if there are only the T and S
amplitudes, the relative phase must depend upon hadron wave functions and
probably reverse with higher excitations, such as $P$-waves or radial
excitations of $S$-waves.

      We may see this already in the case of the $a_1 D$ (eqns B3)
 where unfortunately
the statistics are not good enough to prove anything. 
There are also further tantalising hints in the $B$ system if one
assumes that the $\pi^+ \pi^+ \pi^-$ accompanying the $D^*$, with mass
between $1.0$ and $1.6$GeV, is dominated by the $a_1^+$.
The central values of the data superficially suggest that
here is a final state where the  charged exotic is suppressed relative to
the  all charged 
non-exotic. However, as previously,the errors are such to prevent any meaningful
conclusion.

In both examples involving the $a_1$ production, the T amplitude
depends on the point-like coupling of the $a_1$
 (wave function at the origin) and
an overlap integral of the $B$ and $D$ ground states. The S amplitude 
by contrast depends upon
the point-like coupling of the $D$ (due to the short range
$W$ exchange between the
$c$ and $\bar{d}$) and a p-wave matrix element between the 
$B$ and
$a_1$ ground states.
 Thus the wavefunction overlaps and
the relative phase of the two amplitudes for a final
state with one s-wave meson and one p-wave meson could well be opposite to that
for a final state with two s-wave mesons.
 
      If this is the case one might expect to see a similar effect in charm
decays. Current data on the $a_1 K$ channel are not good enough to decide.
The possibility that the interference sign is channel dependent
may be tested also by data on scalar mesons in $D$ final states, such as
the $K_0(1430)$ and the broad $f_0(1300)$ which are candidate members of the
scalar nonet. Data exist on $D^+ \rightarrow \pi^+ K_0^{*o}(1430)$,
$D^o \rightarrow \pi^+ K_0^{*-}(1430)$;
in the absence of data on $K_0^{*o} \pi^o$ one may
use $f_0(1300) \bar{K}^o$
as the flavour and overall spin structures are the same. 
If one demands that the $S$ amplitude is colour suppressed in magnitude relative 
to the $T$ amplitude, then these channels involving scalar mesons
appear to prefer {\bf destructive} interference.

Further hints that the interference may be destructive in the $D \rightarrow
\pi K^*$ channels comes from their Cabibbo suppressed analogues

$$
br(D^+ \rightarrow \pi^+ \bar{K}^{o*}) \sim (T + S)^2
= 2.2\pm 0.4\%
$$from which if we ignore modifications arising from phase space
and exclusive form factors (which tend to counterbalance\cite{cafe95}),
we may expect

$$
br(D^+ \rightarrow \pi^+ \rho^o) \sim \frac{(T + S)^2}{2} \times sin^2\theta
\sim 0.05\%
$$
(where we have ignored any annihilation or Penguin contribution). This is 
consistent with the data which report $<0.15\%$ for this Cabibbo suppressed
mode and suggests this analysis is robust. Then if we consider
the related Cabibbo suppressed mode

$$
br(D^+ \rightarrow \phi \pi^+) \sim S^2 sin^2\theta = 0.67 \pm 0.08\%
$$
we have a rather clean measure of the strength of the colour suppressed
$S$ diagram. This suggests that the $\rho \pi$ rate is ``small" due to
$T$ and $S$ interference being destructive (or that there is destructive
interference with a $W$ or Penguin topology). The $TS$ destructive 
interference would also suggest that the $\pi^+ \bar{K}^{o*}$ also is
``small" due to destructive interference (which is 
consistent with the analysis of section 2.1 applied to eqns.(4)).

{\it The systematics of constructive and destructive interference appear
from our analysis to be non trivial and channel dependent. The data need to
be sharpened as we have noted if a pattern is to be discerned}.

Elsewhere \cite{cll}
it has been noted that
the $\pi(1.8)$ can contribute to the direct channel in penguin driven
Cabibbo suppressed D decays. The existence of this state is well
established though its internal structure, whether hybrid or radial
excitation, remains to be settled\cite{close94,cp94,paton85}.
 In either case one expects that
there will be a $K$ partner and with a mass $K(\sim 1.9)$. Such a
state will have typical strong decay width of $O(200 MeV)$ and thereby
overlap the D mass; consequently it may be expected to affect the
$0^{-+}$ overall final states in D decays via the $W$-exchange diagram.
 
Analogously, enhancements may be anticipated in the $0^{++}$ overall due to
the presence of the (radial excitation) $K_0(1950); \Gamma \sim 200$, and
possibly Cabibbo suppressed modes by its $f_0$ or $a_0$ partners\cite{buccella}.
 
This is in sharp contrast to the B decays where the required resonances
would be $J^P = 0^-$ or $0^+$ D states around the B mass, namely $\sim 5$GeV.
Unlike the  $K$ and $\pi$ system where the lightest hybrids or prominent
radial excitations are expected around 2Gev and hence in the vicinity of
the (initial state) $D$ meson, the lightest hybrids or prominent
radial excitations of the $D$ with $0^-$ or $0^+$ quantum numbers are
anticipated to be in the $\sim 3.5$GeV region, far below the $5$GeV mass
of the (initial state) $B$ meson.
 
If this is an important source of the D decay $W$ enhancement, one may expect
correlation between those channels and the branching ratios of the
respective $K$ direct channel resonances. In particular it will require the
$I=\frac{1}{2}$ correlation among charged and neutral modes in the
final state.  This appears to be satisfied for $\pi K$ and within
errors for $\pi K^*$; it may also be true
(possibly) for $\rho K^*$ (when one compares the transverse polarization
values for the latter as these are the only two that enable direct comparison
in a single experiment in the PDG\cite{pdg94}). It does not arise for the $a_1 K$
and $\rho K$ where the all neutral modes are much suppressed relative to
their charged counterparts.

\section{Some Sum Rules for Insight from $D$ and $B$ Decay Data}
 
The relations (6) satisfy the sum rule
$$A(\rho^+\bar K^o)  = A(\rho^+K^-)-
{\sqrt 2}\cdot A(\rho^o\bar K^o) \eqno(14) $$
This sum rule is seen to follow 
from general isospin relations. Both sides are
pure $I=3/2$ amplitudes. They are related because the initial state has
$I=1/2$ and the weak interaction operator for these $c \rightarrow s$
transitions is pure $I=1$. In this form the sum rule relates only the exotic
contributions; i.e. colour-favored and colour-suppressed, and projects out
all W exchange and resonance contributions.
Since phases are unknown, this sum rule gives only a triangular
inequality for the experimental branching ratio data. It is interesting
that for this case the data are:
$$
{{\sqrt{\{br(\rho^+K^-)\}}}\over{\sqrt{\{br(\nu e^+K^-)\}}}}=
1.66 \pm 0.11
\pm 3\%
~ ~ ~ ~ ~ ~
{\sqrt 2}\cdot
{{\sqrt{\{br(\rho^o\bar K^o)\}}}\over{\sqrt{\{br(\nu e^+K^-)\}}}}=
0.76 \pm 0.07
\pm 3\%
\eqno(15a) $$
$$
{{\sqrt{\{br(\rho^+\bar K^o)\}}}\over{\sqrt{\{br(\nu e^+\bar K^o)\}}}}=
1.0  \pm 0.18
\pm 6.5\%
\eqno(15b) $$
Then
$$
{{\sqrt{\{br(\rho^+K^-)\}}}\over{\sqrt{\{br(\nu e^+K^-)\}}}}-
{\sqrt 2}\cdot
{{\sqrt{\{br(\rho^o\bar K^o)\}}}\over{\sqrt{\{br(\nu e^+K^-)\}}}}=
0.90 \pm 0.21 \pm 7\%
\eqno(15c) $$
which is within experimental errors of the lower limit of the inequality.
 
A similar approach can be made for $  \rho \bar K^*(892)  $,
$\pi \bar K $ and  $  \pi  \bar K^* $ final states using the data
in section 2. From these we find that the 
sum rules with pions in the final state both have equal
contributions to the sum rule from the two legs of the triangle
for the neutral decay modes, as if the neutral decays were
pure I=1/2. The exact significance of this behavior is unclear without
information on phases.
But the fact that both pion sum rules show similar
behavior and both $\rho$ sum rules show similar behavior and the behavior
of pionic and $\rho$ sum rules are very different from one another may
be significant.
 
On the other hand the final state interactions and possible resonances
are expected to be very different for the even parity (scalar) and
odd parity (pseudoscalar) states since strong interactions conserve
parity.
The two pion sum rules which show similar behavior refer to two states of
opposite parity and the two $\rho$ sum rules which show similar behavior
probably also refer to states of opposite parity. In the vector-vector
case both parities are present but two of the three helicity amplitudes
have even parity whereas the single vector-pseudoscalar amplitude has odd
parity.
 
{\it It will be interesting to see if these results hold up under improved 
statistics and, if so, a challenge to explain them}

 
The analogous sum rule for $B$ decays is more convenienty rearranged to the form

$$A(\rho^+\bar D^o) -  A(\rho^+D^-) = -
{\sqrt 2}\cdot A(\rho^o\bar D^o)
\eqno(B14) $$

In this form the sum rule is seen to cancel the $T$ 
contribution on the LHS and to give two expressions for the combination
$S-W$, which we have seen from the data to be small. For this case, the
sum rule provides the same information as the relation (B11). 
 
The relevant data are:
$$
{{\sqrt{\{br(\rho^+D^-)\}}}\over{\sqrt{\{br(\nu e^+D^-)\}}}}=
0.64 \pm 0.05
\pm 13\%
~ ~ ~ ~ ~ ~
{\sqrt 2}\cdot
{{\sqrt{\{br(\rho^o\bar D^o)\}}}\over{\sqrt{\{br(\nu e^+D^-)\}}}} < 0.28
\eqno(B15a) $$
$$
{{\sqrt{\{br(\rho^+\bar D^o)\}}}\over{\sqrt{\{br(\nu e^+\bar D^o)\}}}}=
0.91 \pm 0.06
\pm 22\%
\eqno(B15b) $$
The upper limit on the RHS is seen to be very near to the lower bound on the
LHS. Thus better data will be able to determine the relative phase of the 
contributing amplitudes, defined by the angle $\theta$ in eq. (11).

A similar approach gives the analogous sum rule for the
$  \rho \bar D^*  $ 
$\pi \bar D $ and  $  \pi  \bar D^* $ final states.
 
In the $D$ decays to $\pi K(^*)$ the data were suggestive that the
neutral decays were dominated by $I=1/2$. This is not the case for
$B^o \rightarrow \pi D$ though this is not ruled out for the
$B^o \rightarrow \pi D^*$.
 
\section{Conclusions - Possible Implications for CP Searches}
 
We have shown interesting systematics in exclusive $D$ and $B$ decays which
warrant future experimental investigation and theoretical analysis. 

One example of possible new systematics would be CP-exotic states which
like flavour-exotic states also cannot arise in the quark-antiquark system and 
cannot be enhanced by $q\bar{q}$ resonances.
However, such states  may exist in a quark-antiquark-gluon configuration, which
can be produced by a W-exchange diagram. There might also be as yet unknown
hybrid $q \bar q G$ states in this region with CP-exotic quantum numbers.
Therefore it is of interest to look for such final states.

We now note that a better understanding of decay systematics can prove useful 
in guiding the choice of useful candidate decay modes for CP-violation studies.
Many proposed searches for CP violation focus on producing $\bar B B$ pairs at
the $\Upsilon (4S)$ and observing a lepton asymmetry in one decay when the other
is observed to decay into a CP eigenstate like $\psi K_S$. Unfortunately there
are not many known unambiguous CP eigenstates. Many final states like $\rho \pi$
have several partial waves with opposite CP eigenvalues. Such states can be used
in CP violation experiments only if the two partial waves 
(and thereby the $CP = \pm 1$ combinations) have been separated 
by partial wave or isospin analysis. These analyses would be completely 
unnecessary if decay systematics show that only partial waves with a given
CP eigenstate are presnt. If, for example only the odd CP partial waves appear 
in the $3 \pi$ final state, all neutral three-pion states could
be used in CP-violation experiments by analogy with $\psi K_S$ without any 
necessity for the selection of $\rho-\pi$ mass peak and isospin analyses. 
This would occur if decays into CP-exotic partial waves were suppressed. 

There are two $J^{PC}$ values allowed for the weak decay of a spin-zero meson 
into a neutral $3\pi$ state; namely $0^{--}$ and $0^{-+}$. Of these the $0^{--}$ is
CP even and has exotic quantum numbers while the $0^{-+}$ is CP odd and has
normal quantum numbers. 
It is therefore of interest to examine the $3 \pi$ final states in $D$
and $B$ decays by Dalitz plots and partial wave
analyses using charged as well as neutral decays. Preliminary data on
$D_s$ decays into three pions \cite{appel}
suggest dominance by CP-odd partial waves.
 
Since the parity-violating weak interaction leads to final states
that overall can be both scalar and
pseudoscalar, the resonance structures at the $D$ and $D_s$ masses can
be quite different for the states of opposite parity. This could show up as a
systematic difference.
 
Note that nonleptonic enhancement in nonexotic channels as well as large
W-exchange contributions are inconsistent with factorization. It is believed
that factorization is a good approximation at sufficiently high mass.
An interesting open question is whether and where a transition between
nonleptonic enhancement and factorization occurs.
 
We are grateful to J.Appel, C.Damerell, Y. Grossman, M. Sokoloff and S.Stone 
for discussions

\def \ajp#1#2#3{Am.~J.~Phys.~{\bf#1} (#3) #2}
\def \apny#1#2#3{Ann.~Phys.~(N.Y.) {\bf#1} (#3) #2}
\def \app#1#2#3{Acta Phys.~Polonica {\bf#1} (#3) #2 }
\def \arnps#1#2#3{Ann.~Rev.~Nucl.~Part.~Sci.~{\bf#1} (#3) #2}
\def \cmp#1#2#3{Commun.~Math.~Phys.~{\bf#1} (#3) #2}
\def \cmts#1#2#3{Comments on Nucl.~Part.~Phys.~{\bf#1} (#3) #2}
\def \cn{Collaboration}
\def \corn93{{\it Lepton and Photon Interactions:  XVI International Symposium,
Ithaca, NY August 1993}, AIP Conference Proceedings No.~302, ed.~by P. Drell
and D. Rubin (AIP, New York, 1994)}
\def \cp89{{\it CP Violation,} edited by C. Jarlskog (World Scientific,
Singapore, 1989)}
\def \dpff{{\it The Fermilab Meeting -- DPF 92} (7th Meeting of the American
Physical Society Division of Particles and Fields), 10--14 November 1992,
ed. by C. H. Albright \ite~(World Scientific, Singapore, 1993)}
\def \dpf94{DPF 94 Meeting, Albuquerque, NM, Aug.~2--6, 1994}
\def \efi{Enrico Fermi Institute Report No. EFI}
\def \el#1#2#3{Europhys.~Lett.~{\bf#1} (#3) #2}
\def \f79{{\it Proceedings of the 1979 International Symposium on Lepton and
Photon Interactions at High Energies,} Fermilab, August 23-29, 1979, ed.~by
T. B. W. Kirk and H. D. I. Abarbanel (Fermi National Accelerator Laboratory,
Batavia, IL, 1979}
\def \hb87{{\it Proceeding of the 1987 International Symposium on Lepton and
Photon Interactions at High Energies,} Hamburg, 1987, ed.~by W. Bartel
and R. R\"uckl (Nucl. Phys. B, Proc. Suppl., vol. 3) (North-Holland,
Amsterdam, 1988)}
\def \ib{{\it ibid.}~}
\def \ibj#1#2#3{~{\bf#1} (#3) #2}
\def \ichep72{{\it Proceedings of the XVI International Conference on High
Energy Physics}, Chicago and Batavia, Illinois, Sept. 6--13, 1972,
edited by J. D. Jackson, A. Roberts, and R. Donaldson (Fermilab, Batavia,
IL, 1972)}
\def \ijmpa#1#2#3{Int.~J.~Mod.~Phys.~A {\bf#1} (#3) #2}
\def \ite{{\it et al.}}
\def \jmp#1#2#3{J.~Math.~Phys.~{\bf#1} (#3) #2}
\def \jpg#1#2#3{J.~Phys.~G {\bf#1} (#3) #2}
\def \lkl87{{\it Selected Topics in Electroweak Interactions} (Proceedings of
the Second Lake Louise Institute on New Frontiers in Particle Physics, 15--21
February, 1987), edited by J. M. Cameron \ite~(World Scientific, Singapore,
1987)}
\def \ky85{{\it Proceedings of the International Symposium on Lepton and
Photon Interactions at High Energy,} Kyoto, Aug.~19-24, 1985, edited by M.
Konuma and K. Takahashi (Kyoto Univ., Kyoto, 1985)}
\def \mpla#1#2#3{Mod.~Phys.~Lett.~A {\bf#1} (#3) #2}
\def \nc#1#2#3{Nuovo Cim.~{\bf#1} (#3) #2}
\def \np#1#2#3{Nucl.~Phys.~{\bf#1} (#3) #2}
\def \pisma#1#2#3#4{Pis'ma Zh.~Eksp.~Teor.~Fiz.~{\bf#1} (#3) #2[JETP Lett.
{\bf#1} (#3) #4]}
\def \pl#1#2#3{Phys.~Lett.~{\bf#1} (#3) #2}
\def \plb#1#2#3{Phys.~Lett.~B {\bf#1} (#3) #2}
\def \pr#1#2#3{Phys.~Rev.~{\bf#1} (#3) #2}
\def \pra#1#2#3{Phys.~Rev.~A {\bf#1} (#3) #2}
\def \prd#1#2#3{Phys.~Rev.~D {\bf#1} (#3) #2}
\def \prl#1#2#3{Phys.~Rev.~Lett.~{\bf#1} (#3) #2}
\def \prp#1#2#3{Phys.~Rep.~{\bf#1} (#3) #2}
\def \ptp#1#2#3{Prog.~Theor.~Phys.~{\bf#1} (#3) #2}
\def \rmp#1#2#3{Rev.~Mod.~Phys.~{\bf#1} (#3) #2}
\def \rp#1{~~~~~\ldots\ldots{\rm rp~}{#1}~~~~~}
\def \si90{25th International Conference on High Energy Physics, Singapore,
Aug. 2-8, 1990}
\def \slc87{{\it Proceedings of the Salt Lake City Meeting} (Division of
Particles and Fields, American Physical Society, Salt Lake City, Utah, 1987),
ed.~by C. DeTar and J. S. Ball (World Scientific, Singapore, 1987)}
\def \slac89{{\it Proceedings of the XIVth International Symposium on
Lepton and Photon Interactions,} Stanford, California, 1989, edited by M.
Riordan (World Scientific, Singapore, 1990)}
\def \smass82{{\it Proceedings of the 1982 DPF Summer Study on Elementary
Particle Physics and Future Facilities}, Snowmass, Colorado, edited by R.
Donaldson, R. Gustafson, and F. Paige (World Scientific, Singapore, 1982)}
\def \smass90{{\it Research Directions for the Decade} (Proceedings of the
1990 Summer Study on High Energy Physics, June 25 -- July 13, Snowmass,
Colorado), edited by E. L. Berger (World Scientific, Singapore, 1992)}
\def \smassb{{\it Proceedings of the Workshop on $B$ Physics at Hadron
Accelerators}, Snowmass, Colorado, June 21 -- July 2, 1993, edited by
P. McBride and C. S. Mishra (Fermilab report Fermilab-CONF-93/267, 1993)}
\def \stone{{\it B Decays}, edited by S. Stone (World Scientific, Singapore,
1994)}
\def \tasi90{{\it Testing the Standard Model} (Proceedings of the 1990
Theoretical Advanced Study Institute in Elementary Particle Physics, Boulder,
Colorado, 3--27 June, 1990), edited by M. Cveti\v{c} and P. Langacker
(World Scientific, Singapore, 1991)}
\def \yaf#1#2#3#4{Yad.~Fiz.~{\bf#1} (#3) #2 [Sov.~J.~Nucl.~Phys.~{\bf #1} (#3)
#4]}
\def \zhetf#1#2#3#4#5#6{Zh.~Eksp.~Teor.~Fiz.~{\bf #1} (#3) #2 [Sov.~Phys. -
JETP {\bf #4} (#6) #5]}
\def \zpc#1#2#3{Zeit.~Phys.~C {\bf#1} (#3) #2}

\end{document}